# Controllable optical bistability and cooling of a mechanical oscillator in a hybrid optomechanical system

Bijita Sarma and Amarendra K. Sarma*
*Department of Physics, Indian Institute of Technology Guwahati, Guwahati-781039, Assam, India*
*aksarma@iitg.ernet.in

We investigate theoretically the effect of optical feedback from a cavity containing an ultracold two level atomic ensemble, on the bistable behavior shown by mean intracavity optical field in an optomechanical cavity resonator. It turns out that the optical bistability can be controlled by tuning the frequency and power of the driving laser and is largely affected by the presence of the atomic ensemble in the feedback cavity. In essence, our work emphasizes the possibility of realization of a controllable optical switch depending on the hybrid interaction, commanding lower threshold power than a single optomechanical cavity. Further, we study the aspect of optomechanical cooling of the mechanical oscillator in this hybrid system and the time evolution of the mean phonon number indicates ground-state cooling of the oscillator in the unresolved sideband regime.

## I. INTRODUCTION

A photon scattered from an object transfers momentum to the scatterer, thereby applying radiation pressure force on it. Braginsky and his co-workers in their seminal papers [1-3], predicted long ago that the radiation pressure induced by the optical field confined in a cavity resonator can couple the optical and mechanical modes of the cavity. If we consider an optomechanical cavity having a movable end mirror, driven by a strong laser pump, the radiation pressure force applied by the cavity optical field becomes influential enough to set even macroscopic end mirrors into motion. The motion of the mechanical oscillator modulates the length of the cavity and the optical intensity in the cavity gets altered in turn. This type of system shows high nonlinearity between optical field and mechanical motion, acting analogous to a Kerr-medium [4-5]. In recent years, optomechanical systems have drawn tremendous research interest owing to the possibility of implementing these systems in ground state cooling of mesoscopic mechanical oscillator [6-9], entanglement of optical and mechanical modes [10-12], optomechanically induced transparency [13-16], nonclassical state generation [17-19] and quantum state transfer between different modes [20-23], among others.

Currently hybrid optomechanical systems are highly in focus due to the versatility of both optical and mechanical components in coupling to different systems such as spins, cold atoms, superconducting qubits etc. [24-33]. In this work, we consider a hybrid optomechanical system consisting of two cavities, one optomechanical and the other containing an ultracold two level atomic ensemble, coupled by a single pump laser. In particular, we study the bistable behavior shown by the cavity optical field in the optomechanical cavity with and without feedback from the atomic cavity. Optical bistability inside a cavity with finite decay time, arising due to the dynamic backaction induced by radiation pressure has been studied in various optomechanical systems [34-41]. In [42], Chen et al. have shown that the bistable optical lattice potential resulting from the optomechanical coupling in a cavity can be engineered to obtain two stable ground states: superfluid and Mott insulator states for the intracavity ultracold atoms, for a single input beam. Here, we discuss the controllability of the bistable behavior of the mean intracavity optical intensity in the optomechanical cavity depending on the system parameters provided by the feedback cavity, allowable under possible experimental parameters.

Furthermore, we briefly investigate the ground state cooling of the mechanical oscillator, which is a prerequisite for observing quantum effects in optomechanical systems. In recent years, owing to the unique platform provided by optomechanical systems to study fundamental aspects of quantum physics, tremendous research activities are directed towards achieving ground state cavity cooling of micromechanical oscillators. Conventional cavity cooling of the mechanical oscillator requires the condition of resolved-sideband regime, i.e. the cavity mode decay rate should be lower than the mechanical oscillator resonance frequency, $k \ll \omega_m$. However, in practical situations, this condition cannot be satisfied easily. For typical mechanical oscillators of kHz range of vibrational frequencies, this condition poses a serious constraint. To get rid of this constraint, the idea of adding high-Q resonances to optomechanical systems has been suggested [25, 43]. Adding such system with lower decay rates can modify the noise spectrum around the sidebands. This method relies on the effect of quantum noise interference.

This paper is organized as follows. In Section II we describe the total Hamiltonian of the system and derive the quantum Langevin equations for the system operators. Section III is devoted to the analysis of bistable behavior shown by the mean intracavity optical field in the optomechanical cavity. Section IV discusses optomechanical cooling of the mechanical oscillator followed by conclusion of our work in Section V.

## II. THEORY

We consider a hybrid optomechanical system consisting of two cavities A and C as shown schematically in Fig. 1. Cavity C, with both the end mirrors fixed contains an ensemble of ultracold two level atoms. Cavity A consists of one fixed end mirror and another movable end mirror with resonance frequency $\omega_m$, effective mass $m$ and decay rate $\gamma_m$. Cavity A is driven by an intense pump laser of frequency $\omega_L$, which exerts a radiation pressure force on the movable end mirror. The output optical field from the cavity A drives the cavity C, and the output from cavity C is again fed back into cavity A.

The Hamiltonian of the whole system, in a frame rotating with the driving laser frequency $\omega_L$, is given by:

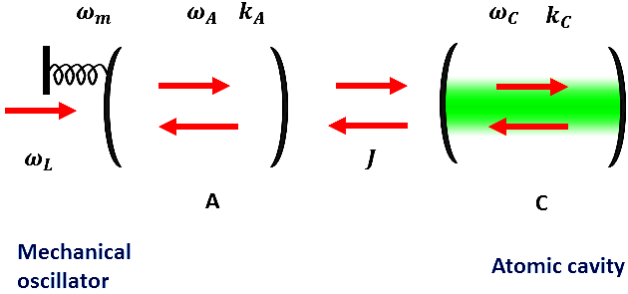

Fig. 1: (Color online) A hybrid optomechanical cavity setup with an optomechanical cavity A and a feedback cavity C containing ultracold atomic ensemble, coupled optically.

$$H = \hbar\Delta_A a^\dagger a + \hbar\Delta_C c^\dagger c + \frac{p^2}{2m} + \frac{1}{2}m\omega_m^2 q^2 + \frac{1}{2}\hbar\Delta_{at}\sigma_3 + \hbar g_{at}(c^\dagger \sigma_{12} + c\sigma_{21}) - \hbar g_{OM} a^\dagger a q + \hbar J(c^\dagger a + a^\dagger c) + i\hbar\varepsilon_A(a^\dagger - a)$$

(1)

where, the first and second terms represent the free energy of the cavity modes in the two cavities A and C respectively. $\Delta_A = \omega_A - \omega_L$ and $\Delta_C = \omega_C - \omega_L$ are the cavity detunings with $\omega_A$ and $\omega_C$ being the corresponding cavity resonance frequencies. The third and fourth terms give the energy of the mechanical oscillator expressed in terms of the position and momentum operators $q$ and $p$ satisfying the commutation relation $[q,p] = i\hbar$. The fifth term is the energy of the two-level atomic ensemble trapped in the cavity C where, $\Delta_{at} = \omega_{at} - \omega_L$, is the detuning of the atomic resonance from the laser drive. $\sigma_{ij}$'s are the atomic operators with $\sigma_{ij} = |i\rangle\langle j|$ and we have denoted $\sigma_{22} - \sigma_{11}$ as $\sigma_3$, where $\sigma_{22}$ and $\sigma_{11}$ are the atomic populations in the excited and ground levels respectively. The sixth term describes the interaction of the atomic ensemble with the optical field in the cavity C, $g_{at}$ being the single atom-photon coupling constant. The seventh term is the optomechanical interaction term, where $g_{OM}$ is the optomechanical coupling constant between the cavity field and the mechanical oscillator in cavity A. The eighth term accounts for the coupling between the two cavities where $J$ is the coupling strength between the two cavities [44-48]. The last term represent the effect of the driving pump laser with frequency $\omega_L$ and amplitude $\varepsilon_A = \sqrt{\frac{2k_A P_{in,A}}{\hbar\omega_L}}$, on the optomechanical cavity, where $P_{in,A}$ is the input laser power.

To study the effect of feedback into cavity A, we first need to analyze the cavity field dynamics in the feedback cavity C. The time evolution of the system operators are given by nonlinear Heisenberg- Langevin equations:

$$\frac{dc}{dt} = -(k_C + i\Delta_C)c - ig_{at}\sigma_{12} - iJa + \sqrt{2k_C}c_{in}(t) \quad (2)$$

$$\frac{d\sigma_{12}}{dt} = -(\gamma_{at} + i\Delta_{at})\sigma_{12} + ig_{at}c\sigma_3 + \sqrt{2\gamma_{at}}c_{in}(t) \quad (3)$$

where, $\gamma_{at}$ is the atomic coherence decay rate and $c_{in}$ is the input vacuum noise operator with zero mean value and nonzero correlation function given by [48]:

$$\langle c_{in}(t) c_{in}^\dagger(t')\rangle = \delta(t-t') \quad (4)$$

Assuming the system operators under mean field approximation and considering $\langle\sigma_{22}\rangle = 0$ and $\langle\sigma_{11}\rangle = N$, i.e. atoms populating only the ground state, the steady state operators are given by:

$$c_S = \frac{-iJa}{k_C + i\Delta_C + \frac{g_{at}^2 N}{\gamma_{at} + i\Delta_{at}}} \quad (5)$$

$$\sigma_{12,S} = \frac{-ig_{at}c_S N}{\gamma_{at} + i\Delta_{at}} \quad (6)$$

Now, defining the dimensionless position and momentum operators $Q$ and $P$ as $Q = \sqrt{\frac{m\omega_m}{\hbar}}\,q$ and $P = \sqrt{\frac{1}{m\hbar\omega_m}}\,p$ for the mechanical oscillator, the equations of motion for the operators for cavity A are given by:

$$\frac{dQ}{dt} = \omega_m P \quad (7)$$

$$\frac{dP}{dt} = \omega_m \chi a^\dagger a - \omega_m Q - \gamma_m P + \xi$$
(8)
(9)

$$\frac{da}{dt} = -i\Delta a - (k_A + \frac{J^2}{k_C + i\Delta_C + \frac{g_{at}^2 N}{\gamma_{at}+i\Delta_{at}}})a + \varepsilon_A + \sqrt{2k_A}a_{in}$$

(9)

where, $\chi = \frac{\omega_A}{\omega_m L}\sqrt{\frac{\hbar}{m\omega_m}}$ is the scaled coupling constant and $\Delta = \Delta_A - \omega_m \chi Q_S$ is the effective detuning in the optomechanical cavity. $a_{in}$ is the input vacuum noise operator for cavity A given by [49]:

$$\langle a_{in}(t)a_{in}^\dagger(t')\rangle = \delta(t-t') \quad (10)$$

$\xi$ is the Brownian noise operator associated with the damping of the mechanical oscillator, with zero mean value and nonzero correlation function given by [50]:

$$\langle \xi(t)\xi(t')\rangle = \frac{1}{2\pi}\frac{\gamma_m}{\omega_m}\int \omega e^{-i\omega(t-t')}[1+\coth(\frac{\hbar\omega}{2k_B T})]d\omega$$

(11)

### III. OPTICAL BISTABILITY

Bistability is a ubiquitous phenomenon observed in many nonlinear systems. The inherent nonlinearity in the equations of motion of our system indicates observation of such effects through optomechanical coupling. Considering that mean values of the system operators can be factorized, one derives the steady state solutions of the equations (7)-(9) as:

$$Q_S = \chi |a_S|^2 \quad (12)$$

$$P_S = 0 \quad (13)$$

$$a_S = \frac{\varepsilon_A}{k_A + \frac{J^2}{k_C + i\Delta_C + \frac{g_{at}^2 N}{\gamma_{at}+i\Delta_{at}}} + i\Delta} \quad (14)$$

Simplifying equation (14), we get the following expression for $|a_S|^2$ that indicates the occurrence of bistable behavior:

$$|a_S|^2[k_{new}^2 + (\Delta_{new} - \omega_m \chi^2 |a_S|^2)^2] = |\varepsilon_A|^2 \quad (15)$$

where, $k_{new} = k_A + \frac{J^2(\gamma_{at}A_1 + \Delta_{at}A_2)}{A_1^2+A_2^2}$ and $\Delta_{new} = \Delta_A + \frac{J^2(\Delta_{at}A_1 - \gamma_{at}A_2)}{A_1^2+A_2^2}$ are the modified optomechanical cavity decay rate and detuning in presence of the atomic cavity; with $A_1 = g_{at}^2 N + k_C\gamma_{at} - \Delta_C\Delta_{at}$ and $A_2 = \Delta_C\gamma_{at} + k_C\Delta_{at}$. Now, to observe bistability, one must have $\frac{\partial|\varepsilon_A|^2}{\partial|a_S|^2} = 0$, which gives, for our system:

$$(k_{new}^2 + \Delta_{new}^2) - 4\Delta_{new}\omega_m\chi^2|a_S|^2 + 3\omega_m^2\chi^4|a_S|^4 = 0$$

(16)

Equation (16) is a quadratic equation in $|a_S|^2$ that will have two distinct roots when the discriminant is positive: $\omega_m^2 \chi^4(\Delta_{new}^2 - 3k_{new}^2) > 0$. This expression clearly shows that for $\chi = 0$, i.e. when there is no optomechanical coupling, bistability disappears. For nonzero $\chi$, the condition for bistability is given by $(\Delta_{new}^2 - 3k_{new}^2) > 0$, or

$$\Delta_A^2 - 3k_A^2 + \frac{J^4}{(A_1^2+A_2^2)^2}(\Delta_{at}A_1 - \gamma_{at}A_2)^2 -$$

$$3\frac{J^4}{(A_1^2+A_2^2)^2}(\gamma_{at}A_1 + \Delta_{at}A_2)^2 - 6k_A\frac{J^2}{(A_1^2+A_2^2)}(\gamma_{at}A_1 +$$

$$\Delta_{at}A_2) + 2\Delta_A \frac{J^2}{(A_1^2+A_2^2)}(\Delta_{at}A_1 - \gamma_{at}A_2) > 0 \quad (17)$$

To analyze the bistability behavior of intracavity optical field in the optomechanical cavity, first we consider the case for $J = 0$, i.e. without coupling to the atomic cavity. In absence of the atomic cavity, the condition in equation

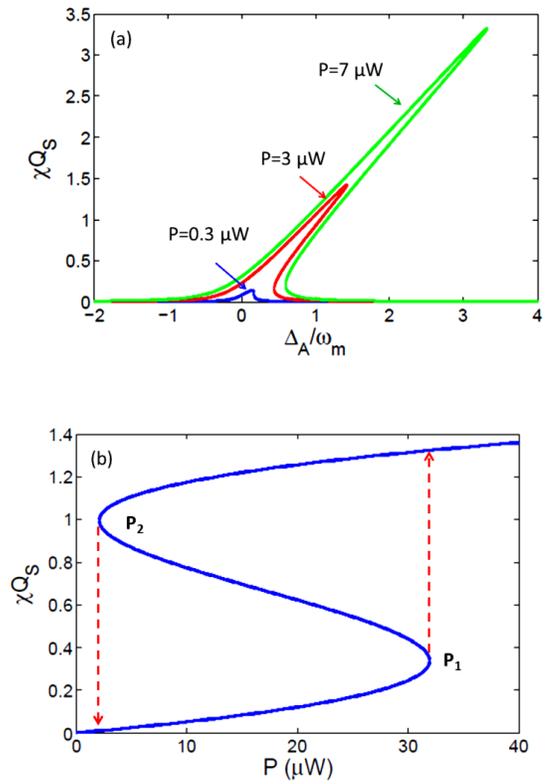

Fig.2: (Color online) Plot of (a) $\chi Q_S$ vs $\frac{\Delta_A}{\omega_m}$ for $P = 0.3\ \mu W$ (blue solid line), $3\ \mu W$ (red solid line) and $7\mu W$ (green solid line), (b) $\chi Q_S$ vs $P$ for optomechanical cavity detuning $\Delta_A = \omega_m$. Other system parameters used are $L = 1mm$, $m = 10\ ng$, $\lambda = 794.98\ nm$, $\omega_m = 10\ MHz$, $k_A = 0.1\ \omega_m$, $Q = 10^7$.

(17) reduces to $\Delta_A^2 - 3k_A^2 > 0$. Fig. 2(a) shows the behavior of intracavity optical intensity in the

optomechanical cavity denoted in terms of $\chi Q_S$ with respect to normalized cavity detuning in the optomechanical cavity, $\Delta_A/\omega_m$. The parameters used are: $L = 1 mm$, $m = 10\ ng$, $\lambda = 794.98\ nm$, $\omega_m = 10\ MHz$, $k_A = 0.1\ \omega_m$, $Q = 10^7$ [51]. The cavity is pumped at the red sideband, $\Delta_A = \omega_m$. For driving laser power $P = 0.3\ \mu W$, the mean intracavity intensity curve is nearly Lorentzian. With increasing power of the driving laser, bistable behavior is seen to occur after crossing a critical value of the input laser power. It is also noted that for higher laser power, bistability occurs at larger cavity detuning.

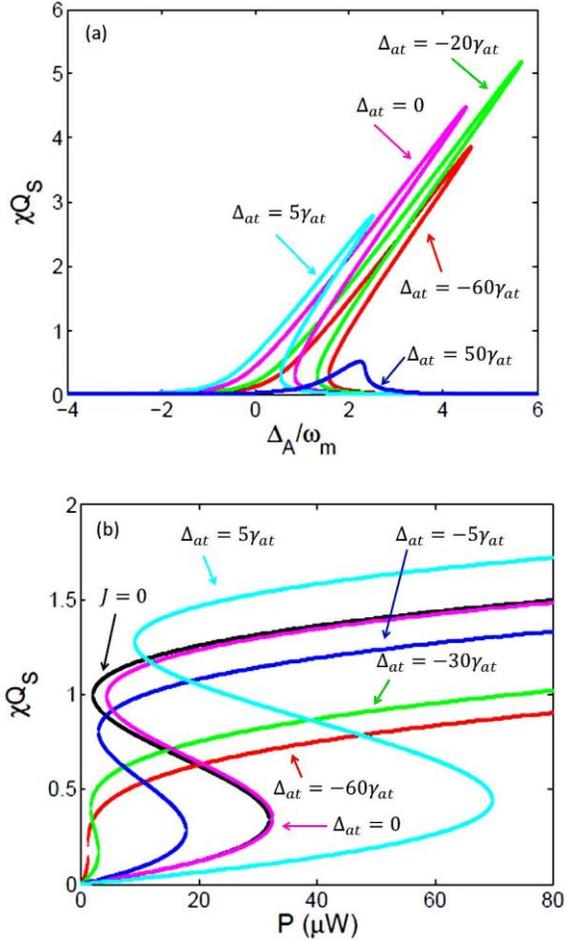

Fig.3: (Color online) (a) Plot of $\chi Q_S$ vs $\Delta_A/\omega_m$, with driving power $P = 20\ \mu W$, (b) Plot of $\chi Q_S$ vs $P$, with cavity detuning $\Delta_A = \omega_m$; for different values of $\Delta_{at}$. Other parameters used are: $J = \omega_m$, $g_{at} = 2\pi\ kHz$, $k_C = 0.1\omega_m$, $N = 10^8$, $\gamma_{at} = 2\pi \times 2.875\ MHz$, $\Delta_C = \omega_m$, others same as in Fig. 2.

Fig. 2(b) exhibits the hysteresis curve for the mean intracavity intensity with respect to varying input power, without feedback from the atomic cavity. This curve clearly indicates the bistable behavior of the intracavity photon intensity. If we start scanning from a low driving power and gradually increase the driving laser power, the intracavity intensity initially follows the lower stable branch. When it reaches the first bistable point $P_1$, it jumps to the upper stable branch and continues to follow that branch for further increasing laser power. Now if we start decreasing the input laser power, the intracavity intensity is observed to decrease following the upper stable branch at first; however when it reaches the second bistable point $P_2$, it will jump down to the lower stable branch and continue to decrease along that branch for further decrease in the input laser power.

Now we proceed to study the bistability behavior of the intracavity optical field in the optomechanical cavity in presence of the atomic cavity. The effect of the presence of the atomic ensemble is shown in Fig. 3. The parameters considered are: $g_{at} = 2\pi\ kHz$, $k_C = 0.1\omega_m$, $N = 10^8$, $\gamma_{at} = 2\pi \times 2.875\ MHz$, $\Delta_C = \omega_m$ and $J = \omega_m$ [31, 43, 46]. Fig. 3(a) shows the bistable behavior shown by the intracavity optical intensity with respect to the detuning in the optomechanical cavity for different values of $\Delta_{at}$ and Fig. 3(b) shows the hysteresis curves. From Fig. 3(a) it is clear that the bistable behavior is dependent on the atom-field detuning $\Delta_{at}$. We get different operating frequency range for bistability for the same input power if we have different atom-field detunings. As can be seen from Fig. 3(a), for higher atom-field detuning, the intracavity optical field curve is nearly Lorentzian. Therefore for higher values of atom-field detuning bistability vanishes. It should be noted that the mean intracavity intensity is highly system specific. In order to satisfy the condition for bistability, the contribution from all the terms in the LHS of Eq. (17) should add up to give a positive number. The bistability behavior also depends on the atom-cavity coupling $g_{at}$ and the coupling between the two cavities $J$, as indicated in equation (17). Fig. 3(b) shows the hysteresis curve for the intracavity optical field. It shows that, for increasing value of atom-field detuning, the operating power range for bistability becomes wider. For the experimental parameters considered in Fig. 3(b), it can be trivially shown that for the system to exhibit bistability, $\Delta_{at}$ should be approximately higher than $-60\gamma_{at}$, which is the threshold value for $\Delta_{at}$ for the specific parameters. The threshold power for observing bistability can be calculated to be: $P_{th} = \frac{\hbar\omega_L}{2k_A}|a_S|^2_{th}[k^2_{new} + (\Delta_{new} - \omega_m\chi^2|a_S|^2_{th})]$, where $|a_S|^2_{th} = \left(2\Delta_{new} - \sqrt{\Delta^2_{new} - 3k^2_{new}}\right)/(3\omega_m\chi^2)$ is the intracavity photon number at the threshold power. For $\Delta_{at} = 0$, the threshold power needed for bistability is almost equal to that for the single optomechanical cavity; and if the detuning is increased to more positive value, the threshold value of input power further shows an increase as seen from Fig. 3(b). Significantly, for negative values of $\Delta_{at}$, the threshold input power for the hybrid system appears to be lower than the generic single cavity case, as shown in Fig. 3(b). The threshold power, $P_{th}$ also depends on the decay rate of the feedback cavity $k_C$.

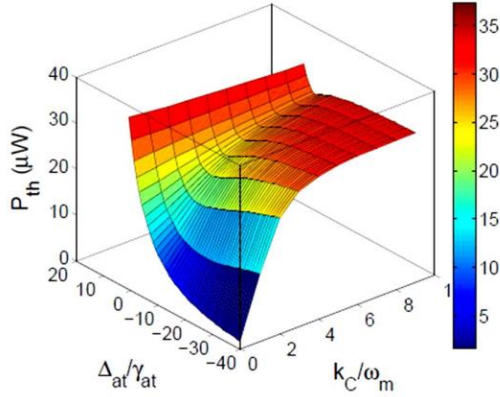

Fig.4: (Color online) Variation of threshold power for bistability as a function of $\Delta_{at}/\gamma_{at}$ and $k_C/\omega_m$.

The variation of $P_{th}$ with respect to $k_C$ and $\Delta_{at}$ is shown in Fig. 4. This shows that for a range of $\Delta_{at}$ and $k_C$, the threshold power is lower than that for the single cavity optomechanical system, which is calculated to be 31.9 $\mu W$ for the experimental parameters considered in this work. Lower value of threshold power is obtained for lower values of $k_C$ as can be seen from Fig. 4. This bistability at low input power is due to the additional feedback field from the atomic cavity. Lower decay rate of the atomic cavity ensures more feedback intensity to the optomechanical cavity. Therefore, coupling a feedback cavity with optimum parameters, to the optomechanical cavity, one can observe bistability for lower input power. Conclusively, we find that the system has the following three externally controllable parameters: power of the single driving laser, cavity-field detuning in the optomechanical cavity and the atom-field detuning in the atomic cavity. The extra controlling parameters provided by the feedback from the atomic cavity modify the condition for bistability. This presents us with more flexibility in switching the intracavity intensity in the optomechanical cavity, between the two stable branches.

## IV. HYBRID OPTOMECHANICAL COOLING

In the previous section, we studied one of the most important consequences of radiation pressure backaction in the hybrid system. Here, we intend to explore cooling of the mechanical mirror that is another celebrated consequence of radiation pressure force. The Hamiltonian of the system can be expressed in a linearized form as

$$H_L = \hbar\Delta a^\dagger a + \hbar\Delta_C c^\dagger c + \hbar\omega_m b^\dagger b + \frac{1}{2}\hbar\Delta_{at}\sigma_3 + \hbar g_{at}(c^\dagger\sigma_{12} + c\sigma_{21}) - \hbar G(a^\dagger + a)(b^\dagger + b) + \hbar J(c^\dagger a + a^\dagger c) \quad (18)$$

Here, $G = g_{OM}\bar{a}$ is the mean intracavity field-enhanced coupling strength and $b$ is the annihilation operator for the mechanical oscillator. The quantum master equation of the system reads

$$\dot\rho = \frac{i}{\hbar}[\rho, H_L] + \frac{k_A}{2}(2a\rho a^\dagger - a^\dagger a\rho - \rho a^\dagger a) + \frac{k_C}{2}(2c\rho c^\dagger - c^\dagger c\rho - \rho c^\dagger c) + \frac{\gamma_{at}}{2}(2\sigma_{12}\rho\sigma_{21} - \sigma_{21}\sigma_{12}\rho - \rho\sigma_{21}\sigma_{12}) + \frac{\gamma_m}{2}(n_{th}+1)(2b\rho b^\dagger - b^\dagger b\rho - \rho b^\dagger b) + \frac{\gamma_m}{2}n_{th}(2b^\dagger\rho b - bb^\dagger\rho - \rho bb^\dagger) \quad (19)$$

Using the covariance approach, we can find out the time evolution of the mean phonon number $n_b(t) = b^\dagger b\,(t)$ [52]. For this, we need to find out the mean values of all the time-dependent second-order moments: $\langle a^\dagger a\rangle$, $\langle a^\dagger b\rangle$, $\langle a^\dagger c\rangle$, $\langle a^\dagger\sigma_{12}\rangle$, $\langle b^\dagger b\rangle$, $\langle b^\dagger c\rangle$, $\langle b^\dagger\sigma_{12}\rangle$, $\langle c^\dagger c\rangle$, $\langle c^\dagger\sigma_{12}\rangle$, $\langle a^2\rangle$, $\langle ab\rangle$, $\langle ac\rangle$, $\langle a\sigma_{12}\rangle$, $\langle b^2\rangle$, $\langle bc\rangle$, $\langle b\sigma_{12}\rangle$, $\langle c^2\rangle$ and $\langle c\sigma_{12}\rangle$. These are determined by solving a linear system of differential equations $\partial_t\langle\hat o_i\hat o_j\rangle = Tr(\dot\rho\hat o_i\hat o_j) = \sum_{m,n}\mu_{m,n}\langle\hat o_m\hat o_n\rangle$ where, $\hat o_i$, $\hat o_j$, $\hat o_m$, $\hat o_n$ are one of the operators: $a^\dagger$, $b^\dagger$, $c^\dagger$, $a$, $b$, $c$ and $\sigma_{12}$. $\mu_{m,n}$ are the corresponding coefficients.

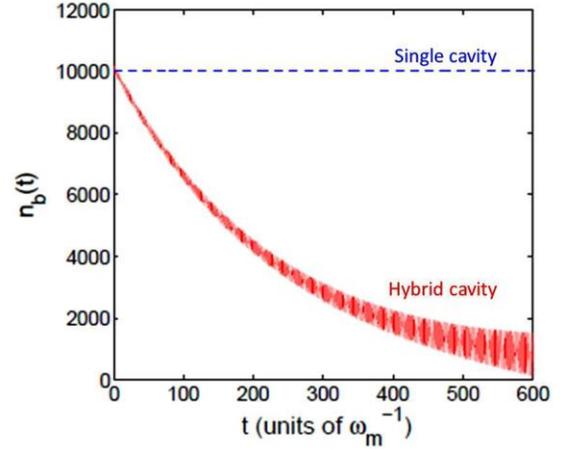

Fig.5: (Color online) Time evolution of the mean phonon number in the mechanical oscillator. The red curve shows the mean phonon number in case of the hybrid cavity with the parameters considered: $\frac{\gamma_m}{\omega_m} = 10^{-5}$, $\frac{k_A}{\omega_m} = 10^2$, $\frac{k_C}{\omega_m} = 1$, $\frac{\gamma_{at}}{\omega_m} = 10^3$, $\frac{g_{at}}{\omega_m} = 0.1$, $\Delta = \omega_m$, $\Delta_C = \omega_m$, $\Delta_{at} = 100\omega_m$ $J = 200\omega_m$ and $G = 50\omega_m$. The single-cavity case (blue-dashed line) with $G = 0.1\omega_m$ is plotted for comparison.

Fig. 5 shows the time evolution of the mean phonon number solved numerically. Initially the phonon number in the mechanical oscillator is equal to the environmental phonon number that is considered to be $10^4$. All other second order moments are initially zero. Other parameters considered in Fig. 5 are: $\frac{\gamma_m}{\omega_m} = 10^{-5}$, $\frac{k_A}{\omega_m} = 10^2$, $\frac{k_C}{\omega_m} = 1$, $\frac{\gamma_{at}}{\omega_m} = 10^3$, $\frac{g_{at}}{\omega_m} = 0.1$, $\Delta = \omega_m$, $\Delta_C = \omega_m$, $\Delta_{at} = 100\omega_m$ and $J = 200\omega_m$, $G = 50\omega_m$. The plot shows the reducing behavior of the mean phonon number with time. On the other hand, the single-cavity case does not show any cooling behavior of the mechanical oscillator as can be seen from Fig. 5. This is clearly an advantage of such hybrid systems that cooling of the mechanical oscillator occurs even in the unresolved-sideband regime for which cooling is not possible in a generic single-cavity optomechanical system.

## V. CONCLUSION

In conclusion, we have studied a hybrid system consisting of an optomechanical cavity and another cavity containing ultracold two level atomic ensemble serving as a feedback to the first cavity; with a special emphasize on bistability shown by the optical field in the optomechanical cavity, emerging due to the effect of radiation pressure force. It turns out that the bistable behavior of the intracavity field in the optomechanical cavity can be controlled by tuning the power of the single driving laser as well as by changing the frequency of the driving laser that can control the cavity-field detuning in the optomechanical cavity and the atom-field detuning in the atomic cavity. This allows more flexibility in controlling bistability compared to the single cavity optomechanical system. In addition, by coupling the atomic cavity with the optomechanical cavity, one can obtain bistability for much lower threshold power compared to the generic optomechanical cavity system. Further, we have studied the aspect of optomechanical cooling of the mechanical oscillator in this system. Ground state cooling of the oscillator is possible in the unresolved sideband regime for large value of cavity decay rate $k_A$.

[1] V. B. Braginsky and A. B. Manukin, Sov. Phys. JETP **25**, 653 (1967).
[2] V. B. Braginsky and V. S. Nazarenko, Sov. Phys. JETP **30**, 770 (1970).
[3] V. B. Braginsky, A. B. Manukin, and M. Yu. Tikhonov, Sov. Phys. JETP **31**, 821 (1970).
[4] C. Fabre, M. Pinard, S. Bourzeix, A. Heidmann, E. Giacobino, and S. Reynaud, Phys. Rev. A **49,** 1337 (1994).
[5] S. Aldana, C. Bruder, and A. Nunnenkamp, Phys. Rev. A **88**, 043826 (2013).
[6] F. Marquardt, J. P. Chen, A. A. Clerk, and S. M. Girvin, Phys. Rev. Lett. **99**, 093902 (2007).
[7] **J**. D. Teufel, T. Donner, D. Li, J. W. Harlow, M. S. Allman, K. Cicak, A. J. Sirois, J. D. Whittaker, K.W. Lehnert, and R.W. Simmonds, Nature (London) **475**, 359 (2011).
[8] J. Chan, T. P. Alegre, A. H. Safavi-Naeini, J. T. Hill, A. Krause, S. Groeblacher, M.Aspelmeyer, and. Painter, Nature (London) **478**, 89 (2011).
[9] A. H. Safavi-Naeini, J. Chan, J. T. Hill, T. P. Mayer Alegre, A. Krause, and O. Painter, Phys. Rev. Lett. **108**, 033602 (2012).
[10] D. Vitali, S. Gigan, A. Ferreira, H. R. Böhm, P. Tombesi, A. Guerreiro, V. Vedral, A. Zeilinger, and M. Aspelmeyer, Phys. Rev. Lett. **98**, 030405 (2007).
[11] C. Genes, A. Mari, P. Tombesi, and D. Vitali, Phys. Rev. A **78**, 032316 (2008).
[12] Y. -D. Wang and A. A. Clerk, Phys. Rev. Lett. **110**, 253601 (2013).
[13] G. S. Agarwal and S. Huang, Phys. Rev. A **81**, 041803 (2010).
[14] S. Weis, R. Riviere, S. Deleglise, E. Gavartin, O. Arcizet, A. Schliesser, and T. J. Kippenberg, Science **330**, 1520 (2010).
[15] A.H. Safavi-Naeini, T. P.Mayer Alegre, J.Chan, M. Eichenfield, M.Winger, Q. Lin, J. T. Hill, D. E. Chang, and O. Painter, Nature (London) **472**, 69 (2011).
[16] H. Wang, X. Gu, Y. Liu, A. Miranowicz, and F. Nori, Phys. Rev. A **90**, 023817 (2014).
[17] S. Mancini, V. I. Man'ko and P. Tombesi, Phys. Rev. A, **55**, 3042 (1997).
[18] S. Bose, K. Jacobs, and P. L. Knight, phys. Rev. A **56,** 4175 (1997).
[19] M. Paternostro, phys. Rev. Lett. **106**, 183601 (2011).
[20] L. Tian and H. L. Wang, Phys. Rev. A **82**, 053806 (2010).
[21] L. Tian, Phys. Rev. Lett. **108**, 153604 (2012).
[22] C. Dong, V. Fiore, M. C. Kuzyk, H. Wang, Science **338**, 1609 (2012).
[23] T. A. Palomaki, J. W. Harlow, J. D. Teufel, R. W. Simmonds and K. W. Lehnert, Nature (London) **495**, 210 (2013).
[24] M. Wallquist, K. Hammerer, P. Rabl, M. Lukin, and P. Zoller, Phys. Scr. T**137**, 014001 (2009).
[25] C. Genes, H. Ritsch and D. Vitali, Phys. Rev. A **80**, 061803(R) (2009).
[26] D. E. Chang, C. A. Regal, S. B. Papp, D. J. Wilson, J. Ye, O. Painter, H. J. Kimble, and P. Zoller, PNAS **107**, 1005 (2010).
[27] P. F. Barker, Phys. Rev. Lett. **105**, 073002 (2010).
[28] K. Stannigel, P. Rabl, A. S. Sørensen, P. Zoller, and M. D. Lukin, Phys. Rev. Lett. **105**, 220501 (2010).
[29] S. Camerer, M. Korppi, A. Jöckel, D. Hunger, T.W.Hänsch, and P. Treutlein, Phys. Rev. Lett. **107**, 223001 (2011).
[30] B. Vogell, K. Stannigel, P. Zoller, K. Hammerer, M. T. Rakher, M. Korppi, A. Jöckel and P. Treutlein, Phys. Rev. A **87**, 023816 (2013).
[31] F. Bariani, S. Singh, L.F. Buchmann, M. Vengalattore and P. Meystre, Phys. Rev. A **90**, 033838 (2014).

[32] A. Dantan, B. Nair, G. Pupilo and C. Genes, Phys. Rev. A **90**, 033820 (2014).
[33] M. Aspelmeyer, T. J. Kippenberg, and F.Marquardt, Rev. Mod. Phys. **86**, 1391 (2014).
[34] M. Vengalattore, M. Hafezi, M. D. Lukin, and M. Prentiss, Phys. Rev. Lett. **101**, 063901 (2008).
[35] F. Brennecke, S. Ritter, T. Donner and T. Esslinger, Science **322**, 235 (2008).
[36] T. P. Purdy, D.W. C. Brooks, T. Botter, N. Brahms, Z. -Y. Ma and D. M. Stamper-Kurn, Phys. Rev. Lett. **105**, 133602 (2010).
[37] R. Ghobadi, A. R. Bahrampur and C. Simon, Phys. Rev. A **84**, 033846 (2011).
[38] E. A. Sete and H. Eleuch, Phys. Rev. A **85**, 043824 (2012).
[39] A. Dalafi, M. H. Naderi, M. Soltanolkotabi and S. Barzanjeh, J. Phys. B: At. Mol. Opt. Phys. **46**, 235502 (2013).
[40] C. Jiang, H. Liu, Y. Cui, X. Li, G. Chen, and X. Shuai, Phys. Rev. A **88**, 055801 (2013).
[41] Y. He, Phys. Rev. A **91**, 013827 (2015).
[42] W. Chen, K. Zhang, D. S. Goldbaum, M. Bhattacharya and P. Meystre, Phys. Rev. A **80**, 011801(R) (2009).
[43] Y. Liu, Y. Xiao, X. Luan, Q. Gong, and C. W. Wong, Phys. Rev. A **91**, 033818 (2015).
[44] Q. Xu, S. Sandhu, M. L. Povinelli, J. Shakya, S.Fan and M. Lipson, Phy. Rev. Lett. **96**, 123901 (2006).
[45] J. Cho, D. G. Angelakis and S. Bose, Phys. Rev. A **78**, 022323 (2008).
[46] Y. Sato, Y. Tanaka, J. Upham, Y. Takahashi, T. Asano and S. Noda, Nat. Photon. **6**, 56 (2012).
[47] C. Zheng, X. Jiang, S. Hua, L. Chang, G. Li, H. Fan and M. Xiao, Opt. Express **20**, 18319 (2012).
[48] B. Peng, S. K. Özdemir, F. Lei, F. Monifi, M. Gianfreda, G. L. Long, S. Fan, F. Nori, C. M. Bender and L. Yang, Nat. Phys.**10**, 394 (2014).
[49] C.W. Gardiner and P. Zoller, *Quantum Noise* (Springer-Verlag, Berlin, 1991).
[50] V. Giovannetti and D. Vitali, Phys. Rev. A **63**, 023812 (2001).
[51] S. Chakram, Y. S. Patil, L. Chang and M. Vengalattore, Phys. Rev. Lett. **112,** 127201 (2014).
[52] Y. Liu, Y. Xiao, X. Luan and C. W. Wong, Phys. Rev. Lett. **110**, 153606 (2013).